\documentclass[conference]{IEEEtran}

\usepackage[cmex10]{amsmath}
\usepackage{amsthm, amssymb}
\usepackage{algorithm}
\usepackage{listings}
\usepackage{tikz}
\usepackage{color}
\usepackage[T1]{fontenc}
\usepackage{graphicx}


\newcounter{parentnumber}

\newtheorem{theorem}{Theorem}
\newtheorem{lemma}{Lemma}
\newtheorem{cor}{Corollary}

\IEEEoverridecommandlockouts

\newcommand{\floor}[1]{\lfloor{#1}\rfloor}

\newtheorem{exam}{Example$\!$}

\newtheorem{remrk}{Remark$\!$}

\begin{document}
\pagenumbering{gobble} 
\sloppy

\usetikzlibrary{shapes}
\newcommand{\alphaVal}{0.01}

\title{On Unique Decoding from Insertions and Deletions} 

\author{Kayvon Mazooji \\ kmazooji1@ucla.edu \\ UCLA, Los Angeles, CA 90095 \thanks{Some of these results were be presented at the IEEE International Symposium on Information Theory (ISIT) 2017.} }






\maketitle

\begin{abstract} 
In this paper, we study how often unique decoding from $t$ insertions or $t$ deletions occurs for error correcting codes.  Insertions and deletions frequently occur in synchronization problems and DNA, a medium which is beginning to be used for long term data storage.

We define natural probabilistic channels that make $t$ insertions or $t$ deletions, and study the probability of unique decoding.  Our most substantial contribution is the derivation of tight upper bounds on the probability of unique decoding for messages passed though these channels. We also consider other aspects of the problem, and derive improved upper bounds for linear codes and VT-codes. 

\end{abstract}  


\section{Introduction}
Codes correcting insertions and deletions have historically been important to problems in synchronization \cite{Ramchandran}.  Recently, such codes have been useful for DNA storage as well \cite{Bruck}.  Researchers have had little luck in finding insertion correcting codes of optimal cardinality.  However, progress has been made on finding upper and lower bounds on the optimal cardinality of such codes \cite{Sala2}, \cite{Negar}, \cite{Negar3}.   The most famous code believed to be optimal is the VT-code \cite{VT}, which corrects a single insertion or deletion \cite{Lev2}. Other codes for insertions are found in \cite{Clayton}, \cite{Abdel-Ghaffar}, \cite{Helberg}, \cite{Schulman}.  List decoding from insertions and deletions is considered in \cite{Antonia}, \cite{Carol}.

In this paper we study the probability of uniquely decoding from insertions and deletions for broad classes of codes.  This problem is particularly relevant to long term storage problems (e.g. DNA storage \cite{Olgica}).  For example, side information may not be available if a codeword is recovered hundreds of years in the future, making  a list of $>1$ decoded codewords insufficient.  It is therefore very desirable for a code to have a reasonable chance of being uniquely decodable beyond the error-correction guarantee, especially if no other properties are compromised.

Consider a channel which makes $t$ insertions into a codeword $c$ and outputs each distinct length $n+t$ received word with equal probability.  The probability of unique decoding for a particular codeword $c$ is equal to the fraction of length $n+t$ received words that are unique to $c.$  We refer to this channel as the \textit{uniform $t$-supersequence channel}.

Now consider the channel where $t$ insertions occur, one at a time.  We refer to the temporal list of insertions as a \textit{$t$-insertion history}.  An insertion is represented by a tuple of the form: (position, element). For the codeword $00,$ an example insertion history is $ [(0, 1),  (0, 0) ], $ giving the received word $0100.$

If for each insertion, the element and position are chosen uniformly at random, we refer to this channel as the \textit{uniform $t$-insertion channel}.   If a particular codeword $c$ in a code $C \subseteq \mathbb{F}_q^n$ is passed through this channel, then the probability that unique decoding occurs is equal to the fraction of $t$-insertion histories that produce a received word unique to $c.$

We also define the analogous channels for $t$ deletion errors in Section II.

Clearly, we can make optimizations involving these probabilities over codes of a particular cardinality, length, and error-correction capability.  While it is not yet clear how to perform such optimizations in a non-exhaustive manner, upper bounds on the probabilities are useful because they help us understand how effective a code can possibly be for uniquely decoding a given number of insertions or deletions, without solving the optimization problems explicitly. The bounds can thus be used as a reference when designing codes, and can be used to prove theorems.  The purpose of focusing on these channels is to provide a framework for understanding what happens when $t$ insertions or deletions occur, regardless of whether the real-life channel always makes $t$ insertions or deletions.

The remainder of the paper is organized as follows.  In Section II, we present the necessary preliminaries.  In Section III we provide tight upper bounds on the probability of unique decoding for both insertion channels, discuss the positivity of the measures, and present improved upper bounds for insertion channels that apply to VT-codes and linear codes. In Section IV, we derive tight upper bounds on the probability of unique decoding for one of the deletion channels, and give improved upper bounds with additional assumptions. In Section V, we make observations about the behavior of VT-codes, and raise open questions.  We conclude the paper in Section VI.  

\section{Preliminaries}

Let $\mathbb{F}_q = \{0,1,\ldots, q-1\}$ be an alphabet containing $q \in \mathbb{N}$ symbols.  We only consider $q \geq 2$ throughout the paper.  $\mathbb{F}_q^n$ is the set of all length $n$ words over the alphabet $\mathbb{F}_q$.  If $a \in \mathbb{F}_q$, we denote the word \[\underbrace{aa\ldots a}_{n \text{ a's}}\]
by $a_n$.  For a sequence $c \in \mathbb{F}_q^n,$ let $c[i]$ be the $i$th element in $c$ where $i \in \{0,1, ...,  n-1\}.$ We use the terms \textit{word} and \textit{sequence} interchangeably.

Let $x \in \mathbb{F}_q^n$.  We define an \textit{insertion} as the addition of an element from $\mathbb{F}_q$ into some position in $x$.  We define a \textit{deletion} as the removal of an element from $x$.  If $t$ insertions occur, the resulting word is referred to as a \textit{$t$-supersequence} of $x$.  Similarly, if $t$ deletions occur, the resulting word is referred to as a \textit{$t$-subsequence} of $x$. A \textit{substring} is a contiguous subsequence of a word.  Given a word, a \textit{run} is a substring $a_n$, such that the potential elements on both sides of $a_n$ are not $a.$  The \textit{Levenshtein distance} $d_L(x, y) $ between two words $ x \in \mathbb{F}_q^{n_1}, \:  y \in \mathbb{F}_q^{n_2},$ is defined as the minimum number of insertions and deletions necessary to transform $x$ into $y$.  Clearly, we have that $d_L(x, y) = d_L(y, x)$.  

The \textit{$t$-insertion ball} of $x$ is the set of all words in $\mathbb{F}_q^{n+t}$ that are formed by inserting $t$ symbols into $x$.  We denote the $t$-insertion ball of $x$ by $I_t(x)$.  It is known that $|I_t(x)| = \sum_{i = 0}^{t}\binom{n+t}{i}(q-1)^i$ for any word $x \in {F}_q^n$. Similarly, the \textit{$t$-deletion} ball of $x$ is the set of all words in $\mathbb{F}_q^{n-t}$ formed by deleting $t$ symbols from $x$.  We denote the $t$-deletion ball of $x$ by $D_t(x)$. Unfortunately, a general formula for $|D_t(x)|$ is not known.  However in the $t=1$ case, we have that $|D_1(x)| = r(x),$ where $r(x)$ is the number of runs in $x$. 

Because $|I_t(x)|$ is independent of the exact length $n$ sequence $x$, we will sometimes use the expression $I_t(n, q)$ to mean the number of sequences in the  $q$-ary $t$-insertion ball of a length $n$ sequence.  

We define a \textit{$t$-insertion correcting code} to be a set of codewords $C \subseteq \mathbb{F}_q^{n}$ such that $I_t(c_1) \cap I_t(c_2) = \emptyset \quad \forall c_1, c_2 \in C, \:  c_1 \neq c_2.$  Similarly, we define a \textit{$t$-deletion correcting code} to be a set of codewords $C \subseteq \mathbb{F}_q^{n}$ such that $D_t(c_1) \cap D_t(c_2) = \emptyset \quad \forall c_1, c_2 \in C, \:  c_1 \neq c_2$.


It was shown that $C$ is a $t$-insertion correcting code if and only if it is a $t$-deletion correcting code. Furthermore, it was also shown that $C$ is a $t$-insertion correcting code if and only if $ d_L(c_1, c_2) > 2t \quad \forall c_1, c_2 \in C, \:  c_1 \neq c_2$.  We denote the \textit{minimum Levenshtein distance} of an insertion/deletion correcting code by $d_{\text{min}}$, and denote the cardinality of the code by $M.$

Varshamov-Tenengolts codes introduced in \cite{VT} are commonly used single-insertion correcting codes defined as the set of all words $x = (x_1, ..., x_n) \in \mathbb{F}_2^n$ such that

\[ \sum_{i=1}^n ix_i \equiv a \: \: ( \! \! \! \! \! \mod n+1) \]  for some $a$ such that  $0 \leq a \leq n.$  

Varshamov-Tenengolts codes are perfect, and are optimal for $n = 1, ..., 9$ when $a=0$.  It is conjectured, though unproven that Varshamov-Tenengolts codes are optimal for all values of $n$ when $a=0$.  We refer to the Varshamov-Tenengolts code of length $n$ with parameter $a$ as $VT_a(n)$.  Some excellent general resources on insertions and deletions can be found in \cite{Sloan}, \cite{PHD_thesis},  and \cite{Mitzenmacher}.

We focus on two distinct insertion channels, namely the {\it uniform $t$-supersequence channel,} which we denote by $\text{USC}_t,$ and the {\it uniform $t$-insertion channel,} which we denote by $\text{UIC}_t.$  These channels are both defined in the introduction.

We consider two deletion channels which make $t$ deletions.  The first is the uniform $t$-deletion channel, denoted by  $\text{UDC}_t.$ This channel makes $t$ sequential deletions, where the deletion at each step is chosen with equal probability.  This is equivalent to saying each $t$-deletion history occurs with equal probability.  Here, a $t$-deletion history is represented as temporal list of deletions of length $t,$ where the $i$th deletion is represented by the index of the element in the length $n-i+1$ word that is deleted at step $i \in \{1, \; ..., t\}.$  For example, if the word $001100$ is affected by the $2$ deletion history $[0, 1], $ the received word would be $0100.$

The second deletion channel is the uniform $t$-subsequence channel, which outputs each distinct $t$-subsequence of a word with equal probability.  We denote this channel by $\text{UBC}_t.$

We define a {\it unique $t$-supersequence} of a codeword $c \in C \subseteq \mathbb{F}_q^n$ as a $t$-supersequence of $c$ that is not a $t$-supersequence of any other codeword in $C.$

We define a {\it unique $t$-subsequence} of a codeword $c \in C \subseteq \mathbb{F}_q^n$ as a $t$-subsequence of $c$ that is not a $t$-subsequence of any other codeword in $C.$

We consider two measures of a code's effectiveness for 
uniquely decoding beyond its error-correction guarantee. Let $f_{K} (c, C)$ be the probability of $c \in C$ being uniquely decodable after being passed through channel $K.$  The first measure is $W_{K}(C) = \min_{c \in C} f_{K} (c, C).$  This measure addresses the worst case distribution on the codeword sent i.e. no matter which codeword is sent, the probability of unique decoding is at least $W_{K}(C).$  The second measure is $U_{K}(C) = \frac{1}{M} \sum_{c \in C} f_{K} (c, C).$  This measure gives the probability of unique decoding if each codeword is sent with equal probability.  

A  $W_K$ optimal code is a code that solves

\[\max_{C: \; \{ n, q, d_{\text{min}}, M \} \; \text{fixed} } W_{K}(C)  .\]

A $U_K$ optimal code is a code that solves

\[\max_{C: \; \{ n, q, d_{\text{min}}, M \} \; \text{fixed} } U_{K}(C)  .\]

Finally, we define $B(n, q, K)$ as a tight upper bound on $f_K(c, C)$ over all codes $C \subseteq \mathbb{F}_q^n$ such that $|C| \geq 2,$ and over all codewords $c \in C.$

Similarly, we define $B_d(n, q, K)$ as a tight upper bound on $f_K(c, C)$ over all codes $C \subseteq \mathbb{F}_q^n$ such that $|C| \geq 2,$ and over all codewords $c \in C$ such that there is a codeword $c' \in C$ where $d_L(c, c') = 2d$ for $ 1 \leq d \leq n.$

\section{Bounds for Insertions}

\subsection{General Bounds}

In this subsection we find $B(n, q, K)$ for $K \in \{ \text{USC}_{t},  \text{UIC}_{t} \}, $  consider the limiting behavior of the upper bounds as $t$ or $n$ increases, and establish the positivity of $U_{K}$ for $K \in \{ \text{USC}_{t},  \text{UIC}_{t}\}$ and all $t.$   Recall that $q \in \mathbb{N},$ $q \geq 2$ throughout the paper.

We begin by recalling a recursion for the intersection cardinality of two insertion balls.  This recursion was discovered by Levenshtein in \cite{Lev3}, where he studied what is now known as Levenshtein's reconstruction problem.  

\begin{lemma}
	\label{levTrick}
	Let $X' \in \mathbb{F}_q^{n+t-k}, Y' \in \mathbb{F}_q^n,$ where $n, t, k$ are positive integers such that $k < n+t.$  Write $X' = aX$ and $Y' = bY$ with $a,b \in \mathbb{F}_q$. Then, if $a=b$,
	\begin{align*} 
	|&I_k(X') \cap I_t(Y')| = \nonumber  \\
	&\quad|I_{k}(X) \cap I_t(Y)| + (q-1) |I_{k-1}(aX) \cap I_{t-1}(aY)|. 
	\end{align*}
	If $a\neq b$, 
	\begin{align*}  
	|&I_k(X') \cap I_t(Y')| = |I_k(X) \cap I_{t-1}(bY)| + \nonumber \\
	&|I_{k-1}(aX) \cap I_t(Y)| + (q-2)|I_{k-1}(aX) \cap I_{t-1}(bY)|. 
	\end{align*}
	
\end{lemma}

We define the {\it minimum intersection cardinality} as \[\ddot{N}^+_q(n_1, n_2, t_1, t_2) = \min_{\substack{X \in F_q^{n_1}, Y \in F_q^{n_2}}} |I_{t_1}(X) \cap I_{t_2}(Y)| \]  where  $n_1,n_2,t_1,t_2 \in \mathbb{N},$ and $t_1 + n_1 = t_2 + n_2.$  Clearly, $\ddot{N}^+_q(n_1, n_2, t_1, t_2) = \ddot{N}^+_q(n_2, n_1, t_2, t_1).$  In Theorem \ref{lem:Main}, we give a closed form for $\ddot{N}^+_q(n_1, n_2, t_1, t_2)$.   Levenshtein derived the maximum intersection analogue of $\ddot{N}^+_q(n_1, n_2, t_1, t_2)$ in his study of sequence reconstruction \cite{Lev3}.

\begin{theorem} \label{lem:Main}

	Let $n_1,n_2,t_1,t_2 \in \mathbb{N}$ and $t_1 + n_1 = t_2 + n_2$.  Then, we have
	
	\begin{align*} \label{eq:N_dots} \ddot{N}^+_q(n_1, n_2, t_1, t_2) =  \sum_{k = n_1}^{t_2} \sum_{i=0}^{k-n_1} \binom{k}{i}(q-2)^i \binom{n_2 + t_2}{k}. \!
	\end{align*}
	
\end{theorem}

To prove Theorem \ref{lem:Main} , we found $I_{t_1}(0_{n_1}) \cap I_{t_2}(1_{n_2})$ to equal the formula in the theorem.  We then proved the formula is the minimum intersection cardinality using an inductive argument on $n_1 + t_1.$  Lemma \ref{levTrick} was used in the inductive step.  The proof is very long, and is given in the appendix.  It should be noted that every pair of sequences at Levenshtein distance $n_1 + n_2$ achieves the minimum intersection cardinality, as given in Lemma \ref{lem:max_pairs_min_intersection}.

	\begin{lemma} \label{lem:max_pairs_min_intersection}
		Let $n_1,n_2,t_1,t_2 \in \mathbb{N}$ and $N = t_1 + n_1 = t_2 + n_2$.  Then for any length $n_1$ sequence $X$ and any length $n_2$ sequence $Y$ such that $d_L(X, Y) = n_1 + n_2$, we have that 
		
		\begin{align*} &|I_{t_1}(X) \cap I_{t_2}Y)| \\
		& = \ddot{N}^+_q(n_1, n_2, t_1, t_2) =  \sum_{k = n_1}^{t_2} \sum_{i=0}^{k-n_1} \binom{k}{i}(q-2)^i \binom{n_2 + t_2}{k}.
		\end{align*} 
	\end{lemma}

In addition, the following corollary can be proved logically, or through an application of the binomial theorem as shown below.
 
\begin{cor}  For $n,t \in \mathbb{N},$ we have that
	\[\ddot{N}^+_q(0, n, n+t, t) =   I_{t}(n, q).\]
\end{cor}

\begin{IEEEproof}
	\begin{align*}
	&\ddot{N}^+_q(0, n, n+t, t) =  \sum_{k = 0}^{t} \sum_{i=0}^{k} \binom{k}{i}(q-2)^i \binom{n + t}{k} \\
	& = \sum_{k = 0}^{t} (q-1)^k \binom{n + t}{k} =  I_{t}(n, q). 
	\end{align*}
	where the second equality follows from the binomial theorem.
\end{IEEEproof}

With Theorem \ref{lem:Main}, we were able to derive $B(n, q, \text{USC}_t).$  Consider any length $n$ code $C$ with cardinality $\geq 2$.  The following theorem gives a tight upper bound on the fraction of unique $t$-supersequences for any $c \in C$.  It is thus an upper bound on the probability of unique decoding for any codeword under the uniform $t$-supersequence channel.

\begin{theorem} \label{theorem:USC_bound}
	
	For $n, t \in N,$ we have that 
	\begin{align*}
	B(n, q, \text{USC}_t) = 1- \Bigg( \frac{ \sum_{k = n}^{t} \sum_{j=0}^{k-n} \binom{k}{j}(q-2)^j \binom{n + t}{k}}{\sum_{i = 0}^{t}\binom{n+t}{i}(q-1)^i} \Bigg).
	\end{align*}

\end{theorem}

\begin{IEEEproof}
	In addition to $c$, there must exist another codeword $c'$ since $|C| \geq 2$. The quantity $|I_t(c) \cap I_t(c')|$ must be greater than or equal to  
	\[\ddot{N}^+_q(n, n, t, t) = \sum_{k = n}^{t} \sum_{j=0}^{k-n} \binom{k}{j}(q-2)^j \binom{n + t}{k} \] as proved in Theorem \ref{lem:Main}.    Thus, the number of sequences unique to $c$ in $I_t(c)$ must be less than or equal to \begin{align*}\sum_{i = 0}^{t}\binom{n+t}{i}(q-1)^i - \sum_{k = n}^{t} \sum_{j=0}^{k-n} \binom{k}{j}(q-2)^j \binom{n + t}{k} \end{align*}
	The upper bound equals $B(n, q, \text{USC}_t)$ because $f_{\text{USC}_t}(0_n, \{0_n, 1_n\}$ achieves the upper bound as proved in Lemma \ref{lem:max_pairs_min_intersection}.
\end{IEEEproof}

In addition to providing an upper bound on the fraction of unique $t$-supersequences for a particular codeword, the formula above serves as a tight upper bound on $W_{\text{USC}_t}(C)$ and $U_{\text{USC}_t}(C).$  

\begin{figure}[H] 
	\centering
	\includegraphics[width=3in]{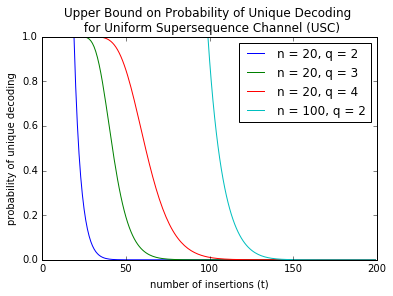}
	\caption{}
	\label{fig:bound_behavior}
	\vspace{-0.1in}
\end{figure}

The upper bound on the fraction of unique $t$-supersequences approaches zero as $t$ goes to infinity for fixed $n$ in the binary case as proved in Lemma \ref{lem:lim_q} and exemplified in Figure \ref{fig:bound_behavior}.  The bound is clearly equal to one as $n$ goes to infinity for fixed $t$ because the minimum intersection cardinality is only positive for  $t\geq n.$ If $t = n + O(n^a)$ for $a <.5,$ our tight upper bound approaches one in the binary case as $n$ goes to infinity.  This is proved in Lemma \ref{lem:lim_n} and is relevant to the questions raised in Section V.

\begin{lemma} \label{lem:lim_q}
	For $n, t \in \mathbb{N},$ we have 
	\begin{align*}
	\lim_{t\to\infty} B(n, 2, \text{USC}_t) = 0.
	\end{align*}
\end{lemma}

\begin{IEEEproof}
	\begin{align*}
	&\lim_{t\to\infty} B(n, 2, \text{USC}_t) \\
	& = \lim_{t\to\infty} {\Bigg( 1-  \frac{ \sum_{k = n}^{t} \sum_{j=0}^{k-n} \binom{k}{j}(q-2)^j \binom{n + t}{k}}{\sum_{i = 0}^{t}\binom{n+t}{i}(q-1)^i} \Bigg) } \\
	&  = 1 - \lim_{t\to\infty} {\frac{ \sum_{k = n}^{t} \binom{n + t}{k}}{\sum_{i = 0}^{t}\binom{n+t}{i}}  } 
	=  0
	\end{align*}
	The first equality follows from plugging in $q=2.$	The second equality follows from the fact that the $n$ extra terms in the denominator are dominated by the other terms as $t$ increases.
	
\end{IEEEproof}

\begin{lemma} \label{lem:lim_n}
	For all values of $n \in \mathbb{N}$ and $a < .5,$ we have 
	\begin{align*}
	\lim_{n\to\infty} B(n, 2, \text{USC}_{n + O(n^a)}) = 1.
	\end{align*}
\end{lemma}

\begin{IEEEproof}
	Let $f: \mathbb{Z} \rightarrow \mathbb{Z}$ be some function. For $q=2$ and letting $t = n + f(n),$ 
	\begin{align*} 
	& \lim_{n\to\infty} B(n, 2, \text{USC}_{n + f(n)}) \\
	& = \lim_{n\to\infty} {\Bigg( 1-  \frac{ \sum_{k = n}^{n+f(n)} \sum_{j=0}^{k-n} \binom{k}{j}(q-2)^j \binom{2n+f(n)}{k}}{\sum_{i = 0}^{n+f(n)}\binom{2n+f(n)}{i}(q-1)^i} \Bigg) }\\
	& = 1- \lim_{n\to\infty} { \frac{ \sum_{k = n}^{n+f(n)} \binom{2n+f(n)}{k}}{\sum_{i = 0}^{n+f(n)}\binom{2n+f(n)}{i}} }  \\
	& \geq  1- \lim_{n\to\infty} \frac{ (f(n)+1) \frac{4^{n + f(n)/2}} {\sqrt{3(n+f(n)/2) +1}} } { {2^{f(n)-1} 4^n}} \\
	& =  1- \lim_{n\to\infty}   \frac{2(f(n)+1)}{\sqrt{3(n+f(n)/2) +1}}. 
	\end{align*}
	
	Letting $f(n) = O(n^a)$ where $a < .5,$ then 
	\[1- \lim_{n\to\infty}   \frac{2(O(n^a)+1)}{\sqrt{3(n+O(n^a)/2) +1}} = 1.\]
	
	The third line follows from plugging in $q=2.$  To obtain the numerator in the fourth line, we observe that $\sum_{k = n}^{n+f(n)} \binom{2n+f(n)}{k} \leq (f(n)+1) \frac{4^{n + f(n)/2}} {\sqrt{3(n+f(n)/2) +1}}$ using the upper bound on the central binomial coefficient in \cite{Kazarinoff}, and the fact the $\binom{2n}{k}$ is maximized at $k = n.$  To obtain the denominator in the fourth line, we observe that $\sum_{i = 0}^{n+w}\binom{2n+f(n)}{i} \geq \sum_{i = 0}^{n+f(n)/2}\binom{2n+f(n)}{i} = \frac{1}{2} 2^{2n+f(n)} = {2^{f(n)-1} 4^n}.$
	
	The proof when $2n+f(n)$ is odd uses the same ideas after an application of Pascal's rule to the binomial coefficients in the third line.	
	
	
\end{IEEEproof}

To improve this upper bound for a code at minimum Levenshtein distance $d_{min}$, we could find  $B_{d_{min}}(n, q, \text{USC}_t).$ This could be approached by finding the minimum insertion-ball intersection cardinality over all pairs of sequences that have Levenshtein distance $\leq d_{min},$ and proving that this measure increases as $d_{min}$ decreases.   Such a formula could be thought of as the dual of the formula derived by Sala et al. in their study of Levenshtein's reconstruction problem \cite{Sala}, which gives the maximum possible insertion-ball intersection cardinality over all sequence pairs with Levenshtein distance $\geq d_{min}.$ The derivation of this formula in conjunction with \cite{Sala} would result in a spectrum of insertion ball intersection cardinalities according to the Levinshetin distance between the sequences in the pair.  It would also be interesting to derive a formula for how many sequence pairs have a particular Levenshtein distance.

To upper bound the probability of unique decoding for the uniform $t$-insertion channel, we find the analogue of the minimum intersection cardinality for insertion histories in Theorem \ref{lem:hist_intersection}.

\begin{theorem} \label{lem:hist_intersection}
	For any distinct sequences $X, Y \in \mathbb{F}_q^n$ with $n, t \in \mathbb{N},$ the fraction of $t$-insertion histories for $X$ giving $t$-supersequences in $I_t(Y)$ is lower bounded by $\sum_{i = 0}^{t-n}\binom{t}{i}(q-1)^i\Pi_{i=1}^t (n+i).$  The lower bound is tight.
\end{theorem}

\begin{IEEEproof}
	The number of insertion histories for $X$ where $Y$ is a subsequence of the insertion pattern is equal to $I_{t-n}(n, q)\Pi_{i=1}^t (n+i) = \sum_{i = 0}^{t-n}\binom{t}{i}(q-1)^i \Pi_{i=1}^t (n+i).$  This is because there are $\Pi_{i=1}^t (n+i)$ ways to choose the positions for an insertion history.  Once the insertion positions are chosen, the insertions clearly appear in some order in the resulting sequence.  So for each history of insertion positions, we can assign a length $t$ sequence to the ordered insertion positions in the resulting sequence.  There are $I_{t-n}(n, q)$ sequences that have $Y$ as a subsequence.  Thus, there are $I_{t-n}(n, q)\Pi_{i=1}^t (n+i)$ total insertion histories, where $Y$ is a subsequence of the insertion pattern.

	So, the number of $t$ insertion histories for $X$ giving $t$-supersequences in $I_t(Y)$ is lower bounded by $\sum_{i = 0}^{t-n}\binom{t}{i}(q-1)^i\Pi_{i=1}^t (n+i).$
	
	The pair $X = 0_n$ and $Y = 1_n$ achieves this upper bound because $0_n$ and $1_n$ have no elements in common, so the only elements in a $t$-supersequence of $X$ that form a copy of $Y,$ must be inserted elements.  By the same logic, any pair at Levenshtein distance $2n$ achieves the upper bound;
\end{IEEEproof}

\begin{theorem} \label{theorem:UIC_bound}
	
	For $n, t \in \mathbb{N},$ we have that 
	\begin{align*}
	B(n, q, \text{UIC}_{t}) = \frac{\sum_{i = t-n+1}^{t}\binom{t}{i}(q-1)^i}{q^t}.
	\end{align*} 
\end{theorem}

\begin{IEEEproof}
	There must be another codeword $c'.$ The number of $t$ insertion histories for $c$ giving $t$-supersequences in $I_t(c')$ is lower bounded by $\sum_{i = 0}^{t-n}\binom{t}{i}(q-1)^i\Pi_{i=1}^t (n+i).$  
	
	There are $q^t \Pi_{i=1}^t (n+i)$ insertion histories total.
	
	Thus, \begin{align*}  & f_{\text{UIC}_t}(c, C) \leq 1 - \frac{\sum_{i = 0}^{t-n}\binom{t}{i}(q-1)^i\Pi_{i=1}^t (n+i)}{q^t \Pi_{i=1}^t (n+i)} \\ 
	&= \frac{\sum_{i = t-n+1}^{t}\binom{t}{i}(q-1)^i\Pi_{i=1}^t (n+i)}{q^t \Pi_{i=1}^t (n+i)} \\
	&= \frac{\sum_{i = t-n+1}^{t}\binom{t}{i}(q-1)^i}{q^t}   \end{align*} where the second line follows from the binomial theorem.
	
	$f_{\text{UIC}_t}(0_n \{0_n, 1_n\})$ achieves the upper bound as shown in the proof of Theorem \ref{lem:hist_intersection}, so the upper bound equals $B(n, q, \text{UIC}_{t}).$
\end{IEEEproof}

\begin{figure}[H] 
	\centering
	\includegraphics[width=3in]{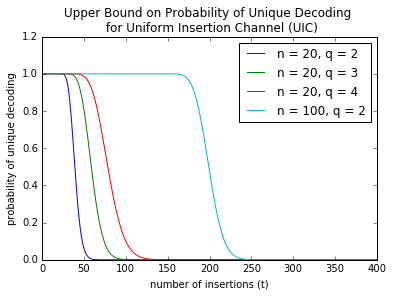}
	\caption{}
	\label{fig:UIC_bound_behavior}
	\vspace{-0.1in}
\end{figure}

$B(n, q, \text{UIC}_{t})$ is also an upper bound on $U_{\text{UIC}_{t}}(C)$ and $W_{\text{UIC}_{t}}(C).$ The following two lemmas give the limiting behavior of the tight upper bound as $t$ and $n$ increase.  The behavior in the lemmas is exemplified in Figure \ref{fig:UIC_bound_behavior}. Interestingly the bound in Figure \ref{fig:UIC_bound_behavior} decays much slower than the bound for $\text{USC}_t$ in Figure \ref{fig:bound_behavior}.

\begin{lemma}  For $n, t \in \mathbb{N}, $ we have 
	$\lim_{t \to \infty} B(n, q, \text{UIC}_{t})  =0.$
\end{lemma}

\begin{IEEEproof} We have  
	\begin{align*}
		\lim_{t \to \infty}  B(n, q, \text{UIC}_{t}) = \lim_{t \to \infty}\frac{\sum_{i = t-n+1}^{t}\binom{t}{i}(q-1)^i}{q^t} = 0.	\end{align*}
	because there are a finite number of terms in the numerator, each of which grows slower than $q^t.$
\end{IEEEproof}

\begin{lemma} \label{lem:UIC_lim_n}  
	For $n \in \mathbb{N}$ and constant $b \in \mathbb{N},$ we have 
	$\lim_{n \to \infty} B(n, q, \text{UIC}_{n+b}) =1.$
\end{lemma}


\begin{IEEEproof}
	\begin{align*} &\lim_{n \to \infty} B(n, q, \text{UIC}_{n+b}) =  \lim_{n \to \infty} \frac{\sum_{i = b+1}^{n+b}\binom{n+b}{i}(q-1)^i}{q^{n+b}} \\ 
	& = \lim_{n \to \infty} 1 - \frac{\sum_{i = 0}^{b}\binom{n+b}{i}(q-1)^i}{q^{n+b}}  \\ 
	& = 1 - \lim_{n \to \infty} \frac{\sum_{i = 0}^{b}\binom{n+b}{i}(q-1)^i}{q^{n+b}} = 1  
	\end{align*}
	The second line follows from the binomial theorem, and the last equality follows because there are a constant number of terms in the numerator that each grow slower than the denominator.
\end{IEEEproof}

In addition to showing upper bounds on the probabilities of unique decoding, we can show that for both insertion channels and every code $C,$ there exists some codeword where the probability of unique decoding is non-zero for all $t.$  This proves the positivity of $U_K(C)$ for all codes $C,$ for both channels of interest.


\begin{lemma} \label{lem:non-neg}
	For every code $C \subseteq \mathbb{F}_q^n,$ and $n \in \mathbb{N}$ there exists a codeword $c \in C$ such that for all $t \in \mathbb{N},$ $c$ has at least one unique $t$-supersequence.
\end{lemma}

\begin{IEEEproof}
	Consider a codeword $c$ with maximal value of $r_{r}(c),$ where $r_{r}(c)$ is the length of the right-most run in $c.$  Suppose that for some $t \in \mathbb{N},$ $c$ did not have at least one unique $t$-supersequence. Let $a$ be the substring of $c$ to the left of $c$'s rightmost run. Consider the supersequence $s$ formed by duplicating the right-most element $t$ times.    Then one or more length $n$ subsequences of $s$ must be codewords in $C$ that are distinct from $c.$    Any such codeword $c'$ must be a length $n$ subsequence of $s.$  $c'$ therefore must be of the form $xy$ where $x$ is a strict subsequence of $a,$ and $y$ is the rightmost element of $c$ repeated $n-|x|$ times. $x$ must be a strict subsequence of $a,$ because if it was not, then $c'$ would be the same as $c.$  Since $x$ is a strict subsequence, $y$ must be of length greater than $ r_{r}(c).$  Thus, $c'$ has a rightmost run of length greater than $ r_{r}(c).$ This is a contradiction.
\end{IEEEproof}

\begin{theorem} \label{theorem:lowerUt}
	For every code $C \subseteq \mathbb{F}_q^n,$ and $n, t \in \mathbb{N},$ we have that $U_K(C)$ is strictly positive for $K \in \{ \text{USC}_{t},  \text{UIC}_{t}\}. $
\end{theorem}

\subsection{Bounds with Additional Assumptions}

In this subsection, we find upper bounds for the probability of unique decoding when additional assumptions are made about the code and the codeword.  All of the results here are applicable to VT-codes of even length with $a=0.$  Recall that VT-codes with $a=0$ have maximal cardinality over all choice of $a,$ and are asymptotically optimal single insertion/deletion correcting codes.


\begin{lemma} \label{lem:all_non_unique_if_0n_and_1n}
	For any binary code $C \subseteq \mathbb{F}_2^n$ with $n \in \mathbb{N},$ such that  $|C| \geq 3$ and $\{0_n, 1_n\} \subset C,$ we have that $f_K(c, C) = 0$ for any $c \notin \{0_n, 1_n\},$ for $ K \in \{ \text{USC}_{t},  \text{UIC}_{t}\}$ and $t \geq n-1, t \in \mathbb{N}.$
\end{lemma}

\begin{IEEEproof}
	For any codeword $c \notin \{0_n, 1_n\},$ consider $I_{n-1}(c).$  Suppose $c$ is composed of $x_1$ $1$'s and $x_0$ $0$'s.  Observe that $1 \leq x_0, x_1 \leq n-1.$
	
	By proving the statement for $t = n-1,$ we prove the statement for $t \geq n-1.$  Consider a sequence $s$ in $I_{n-1}(c),$ and let $s_1$ be the number of ones in $s,$ and $s_0$ be the number of zeros in $s.$ 
	
	If $\geq n - x_1$ ones are inserted, then $s$ is a supersequence of $1_n.$  If $< n - x_1$ ones are inserted, then $> n-1 - (n-x_1) = x_1 - 1$ zeros are inserted, and thus $s_0 > x_0 + x_1 - 1.$  So, $s$ is  a supersequence of $0_n.$
\end{IEEEproof}

\begin{cor}
	For any binary code $C \subseteq \mathbb{F}_2^n$ with $n \in \mathbb{N},$ such that  $|C| \geq 3$ and $\{0_n, 1_n\} \subset C,$ we have that $W_K(C) = 0$ for $ K \in \{ \text{USC}_{t},  \text{UIC}_{t}\}$ and $t \geq n-1, t \in \mathbb{N}.$	
\end{cor}

\begin{cor}
	For any binary code $C \subseteq \mathbb{F}_2^n$  with $n \in \mathbb{N},$ such that  $|C| \geq 3$ and $\{0_n, 1_n\} \subset C,$ we have that $U_K( C) \leq \frac{1}{|C|}$ for $ K \in \{ \text{USC}_{t}\}$ and $t \geq n-1, t \in \mathbb{N}.$	
\end{cor}

\begin{IEEEproof}
	There are $|C| -2$ codewords such that  $f_{ \text{USC}_{t}}(c, C) = 0.$  Therefore, at least one codeword $c \notin \{0_n, 1_n\}$ has all of its $t$-supersequences in $I_t(0_n) \cup I_t(1_n).$  Thus, the number of non-unique $t$-supersequences in $I_t(0_n) \cup I_t(1_n)$ is $\leq 2I_t(n, q)- I_t(n, q) = I_t(n, q).$ 
\end{IEEEproof}

$VT_0(n)$ codes of even $n$ always contain $0_n$ and $1_n.$  So the above results apply.

%

With knowledge of a codeword's weight, we are able to find upper bounds for $\text{UIC}_t$ that drop below $1$ before $t=n.$  These results are also applicable to VT-codes.

\begin{lemma} \label{lem:weight_bound_with_0n_1n}
	For any binary code $C \subseteq \mathbb{F}_2^n$ with $n \in \mathbb{N},$ such that  $|C| \geq 2$ and $\{0_n, 1_n\} \subset C,$ we have that for any codeword $c \in C$ with weight $1 \leq w \leq n-1,$ it follows that $f_{UIC_t}(c, C) \leq 1 - \frac{ \sum_{i=n-w}^t \binom{t}{i} + \sum_{i=w}^t \binom{t}{i} }   {2^t}$ for $1 \leq t \leq n-1, t \in \mathbb{N}.$  
\end{lemma}

\begin{IEEEproof}
	There are $\sum_{i=n-w}^t \binom{t}{i} \Pi_{j=1}^t (n+j)$ $t$-insertion histories for $c$ that give sequences in $I_t(1_n).$  Call this set of insertion histories $H_1.$  
	
	There are $\sum_{i=w}^t \binom{t}{i} \Pi_{j=1}^t (n+j)$ $t$-insertion histories for $c$ that give sequences in $I_t(0_n).$  Call this set of insertion histories $H_2.$
	
	Each history in $H_1$ consist of $\geq n-w$ one insertions, and thus $\leq t - (n-w)$ zero insertions.  For $t \leq n-1,$ this implies that $\leq w-1$ zero insertions are made.  Each history in $H_2$ consists of $\geq w$ zero insertions, and thus, $H_1 \cap H_2 = 0.$  
	
	There are $2^t \Pi_{j=1}^t (n+j)$ insertion histories total, so the result follows.
\end{IEEEproof}

%
%

\begin{lemma} \label{lem:weight_bound_with_0n}
	For any code $C \subseteq \mathbb{F}_q^n$ with $n \in \mathbb{N},$ such that  $|C| \geq 2$ and $0_n \in C,$ we have that for any codeword $c \in C$ with weight $w \geq 1,$ it follows that $f_{UIC_t}(c, C) \leq \frac{ \sum_{i=0}^{w-1} \binom{t}{i} (q-1)^{t-i} }   {q^t}$ for $t \in \mathbb{N}.$  
\end{lemma}

\begin{IEEEproof}
	There are $\sum_{i=0}^{w-1} \binom{t}{i} (q-1)^{t-i}  \Pi_{j=1}^t (n+j)$ $t$-insertion histories for $c$ that give sequences not in $I_t(0_n).$  This is because only $t$-superseqeunces containing $ \leq w-1$ inserted zeros are not in $I_t(0_n).$ For each number of inserted zeros $i,$ the there are $\Pi_{j=1}^t (n+j)$ ways to choose the insertion position history.  Given the $i$ and the insertion position history, there are $\binom{t}{i}$ ways to choose which insertions are zero insertions, and $(q-1)^{t-i}$ ways to choose the elements for the remaining $t-i$ insertions.
	
	There are $q^t \Pi_{j=1}^t (n+j)$ insertion histories total, so the result follows.
\end{IEEEproof}

\begin{figure}[H] 
	\centering
	\includegraphics[width=3in]{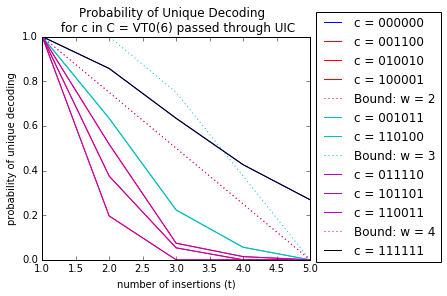}
	\caption{Note: The red curves are underneath the purple curves. The blue curve is underneath the black curve.}
	\label{fig:VT0_6_weight_bound_plot}
\end{figure}

Lemma \ref{lem:weight_bound_with_0n_1n} applies to $VT_0(n)$ of even length. Lemma \ref{lem:weight_bound_with_0n} holds for any linear code since all linear codes contain $0_n.$  Lemma \ref{lem:weight_bound_with_0n} also holds for $VT_0(n)$ of any length.   Provided there is a codeword $c \in C$ of weight $w,$ the corresponding bounds in Lemmas \ref{lem:weight_bound_with_0n_1n} and \ref{lem:weight_bound_with_0n} clearly serve as upper bounds on $W_{\text{UIC}_t}(C).$ Given the codeword weight distribution of a code, Lemmas \ref{lem:weight_bound_with_0n_1n} and \ref{lem:weight_bound_with_0n} can be used to upper bound $U_{\text{UIC}_t}(C).$  In Figure \ref{fig:VT0_6_weight_bound_plot}, we plot the bound in Lemma \ref{lem:weight_bound_with_0n_1n} and the probability of unique decoding for the codewords in $VT_0(6)$ for the uniform insertion channel.  The plot also exemplifies the behavior at $t=n-1$ presented in Lemma \ref{lem:all_non_unique_if_0n_and_1n}.

%
%
%

\section{Bounds for Deletions}

In this section, we consider the uniform $t$-deletion channel $\text{UDC}_t,$ which is defined in Section II.  Specifically, we derive  $B_d(n, q, \text{UDC}_t),$ and derive improved upper bounds with additional assumptions. 

Recall that $\text{UDC}_t$ incurs each $t$-deletion history with equal probability. As an example, if the word $001100$ is affected by the $2$-deletion history $[0, 1], $ the received word would be $0100.$

Clearly, without any assumptions other than $|C| \geq 2,$ the tight upper bound on the probability of unique decoding is $1$ if $t <n,$ and $0$ if $t=n,$ for both the uniform $t$-deletion channel and the uniform $t$-subsequence channel.  This can be seen by taking the code $C = \{0_n, 1_n\}.$

Our main result for this section is the derivation of $B_d(n, q, \text{UDC}_t),$ a tight upper bound on the probability of unique decoding for the uniform $t$-deletion channel, for a codeword $c \in C,$ when there is guaranteed to be a codeword $c' \in C$ such that $d_L(c, c') = 2d$ for $ 1 \leq d \leq n.$ 

To accomplish this, we must first prove a bijection between the set of all $t$-deletion histories to the set of all $t$-deletion patterns.  Here, a $t$-deletion pattern is a temporal list of indices of the original elements removed from the codeword.  For example, if the word $c = 001100$ is affected by the $2$-deletion pattern $[0, 1],$ the received word would be $1100.$  This is because the first element deleted is $c[0], $  and the second element deleted is $c[1].$     

\begin{lemma} 
	There exists a bijection between the set of $t$-deletion patterns and the set of $t$-deletion histories.  
\end{lemma}

\begin{IEEEproof} 
	We will show there is a bijective mapping from the set of $t$-deletion histories to the set of $t$-deletion patterns.  Given a $t$-deletion history, we map it to the resulting $t$-insertion pattern that occurs.  
	
	Any $t$-deletion pattern clearly has a $t$-deletion history that maps to it.  Thus the mapping is onto.  
	
	Now consider the two distinct $t$-deletion histories.  Consider the first position $k$ where they differ.  After the first $k-1$ deletions, the two deletion histories give the same sequence (with same original codeword elements).  Since the $k$th deletion is different, a different original element from the codeword is deleted from the word.  Thus the corresponding deletion patterns differ in the $k$th element.  Thus, the mapping is also  one-to-one, and is thus a bijection. 
\end{IEEEproof}

\begin{figure}
\centering
\includegraphics[width=3in]{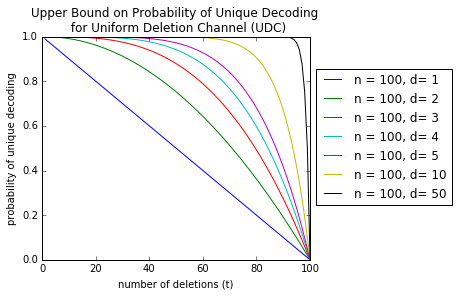}
\caption{}
\label{fig:UDC_upper}
\vspace{-0.1in}
\end{figure}

We now give the central result for the uniform $t$-deletion channel.

\begin{theorem} \label{theorem:UDC_bound}
	 
	For $n, t \in \mathbb{N}$ and $ 1 \leq d \leq n,$ we have that 
	\begin{align*}
		& B_d(n, q, \text{UDC}_t) = \begin{cases} 
		1 & \text{for } 1 \leq t < d \\
		1 - \frac{d! (t-d)! \binom{t}{d}  \binom{n-d}{t-d}(n-t)!}{ n!} & \text{for } d \leq t \leq n
		\end{cases}.
	\end{align*}     
\end{theorem}

\begin{IEEEproof}
	The total number of $t$-insertion histories is equal to $\frac{n!}{(n-t)!}.$ The formula works when $t = n$  because $0! = 1.$
	
	We proceed to work with the equivalent $t$-deletion pattern definition to calculate the numerator in the bound.
	
	Since the Levenshtein distance between $c$ and $c'$ is $2d,$ we know that there exists some set $S$ of $d$ positions in $c,$ such that if we delete the elements at those positions from $c,$ the resulting word will be a $d$-subsequence of $c'.$  We begin by counting the number of ways to select the $d$ steps in the $t$-deletion pattern corresponding to the deletions at indices in $S.$  There are $\binom{t}{d}$ ways to choose these $d$ steps in the deletion pattern.  Fixing those steps in the deletion pattern, there are $d!$ ways to order the deletions.  There are then $\binom{n-d}{t-d}$ ways to choose the remaining indices corresponding to elements in $c$ that are deleted.  Once these are chosen, there are $(t-d)!$ ways to order these deletions. Thus, there are at least $d! (t-d)! \binom{t}{d}  \binom{n-d}{t-d}$ insertion patterns that give subsequences in $D_t(c) \cap D_t(c').$ 

	This bound is tight because the codeword $1_{d} 0_{(n-d)}$ in the code $\{1_{d}0_{(n-d)}, \;  0_n\}$ clearly achieves the bound since there is only one possible set $S$ of size $d.$  This is $S = \{0, \; 1,\;  .., \; d-1\}.$   
\end{IEEEproof}
	
	Interestingly, this bound is alphabet free. The result also serves as an upper bound on $W_{\text{UDC}_t}(C).$  Because the upper bound is tight, the asymptotic analysis of a constant number of extra insertions is of interest.  This is given in the Lemma \ref{lem:UDC_lim_n}.  

\begin{lemma} \label{lem:UDC_lim_n}
	For constant $b \in \mathbb{N},$ we have that  
	\begin{align*} \lim_{n->\infty} B_d(n, q, \text{UDC}_{d+b}) = 1.	\end{align*}
\end{lemma}

\begin{IEEEproof}
	We have that 
	\begin{align*}
		&\lim_{n->\infty} B_d(n, q, \text{UDC}_{d+b}) \\
		& = \lim_{n->\infty} 1 - \frac{d! b! \binom{d+b}{d}  \binom{n-d}{b}(n-d-b)!}{ n!} \\
		& =  1 - \lim_{n->\infty} \frac{d! b! \binom{d+b}{d}  (n-d)! (n-d-b)!}{ (n-d-b)! b! n!} \\
		& =  1 - \lim_{n->\infty} \frac{d! \binom{d+b}{d}  (n-d)!}{ n!} \\
		& =  1 - \lim_{n->\infty} \frac{d! \binom{d+b}{d}  }{ \Pi_{i = 0}^{d-1} (n-i)} = 1.
	\end{align*}
\end{IEEEproof}

	One of the difficulties in finding an analogous upper bound for the uniform $t$-subsequence channel lies in the fact that there is no general formula for the size of a $t$-deletion ball.

	In contrast to the result for insertions in Lemma \ref{lem:non-neg}, there exist codes where all codewords have 0 unique $t$-subsequences for $t \geq \floor{ \frac{d_{min}}{2}}+1$ deletions e.g. $VT_0(6).$  Thus, the average probability of unique decoding for the uniform deletion channel and the uniform subsequence channel is not always positive. This is an example of how the unique decoding of extra deletions differs from the unique decoding of extra insertions. 
	
	Finally, we can find upper bounds on the probability of unique decoding for $\text{UDC}_t$ which make additional assumptions about the code.  The following is the analogue of Lemma  \ref{lem:weight_bound_with_0n_1n} for deletions.

\begin{lemma} 
	For any binary code $C \subseteq \mathbb{F}_2^n$ with $n \in \mathbb{N},$ such that  $|C| \geq 2$ and $\{0_n, 1_n\} \subset C,$ we have that for any codeword $c \in C$ with weight $1 \leq w \leq n-1,$ it follows that 
	\begin{align*} & f_{\text{UDC}_t}(c, C) \\
	&\leq 1 - A_{w}^n \bigg(t, \;  \frac{(n-t)! w! (t-w)! \binom{t}{w}  \binom{n-w}{t-w}}{ n!}\bigg) \\
	& - A_{n-w}^n \bigg( t, \; \frac{ (n-t)! (n-w)! (t-n+w)! \binom{t}{n-w}  \binom{w}{t-n+w}}{ n!} \bigg)
	\end{align*} 
	for $1 \leq t < n, t \in \mathbb{N},$ and 	\begin{align*} f_{\text{UDC}_t}(c, C) = 0
	\end{align*}
	for $t = n,$ where  \begin{align*} A_x^n(t, e) =  \begin{cases}  0 & \text{for } 1 \leq t < x
\\	e & \text{for } x \leq t < n
	\end{cases}.
	\end{align*}  
\end{lemma}

\begin{IEEEproof}
	The number of $t$-deletion histories giving sequences in $|D_t(c)\cap D_t(0_n)|$ is equal to $w! (t-w)! \binom{t}{w}  \binom{n-w}{t-w}$ by the logic used in the Theorem \ref{theorem:UDC_bound}.  Similarly, the number of $t$-deletion histories giving sequences in $|D_t(c)\cap D_t(1_n)|$ is equal to $ (n-w)! (t-n+w)! \binom{t}{n-w}  \binom{w}{t-n+w}.$  The first set of insertion histories results in sequences of all 0's. The second set of insertion histories results in sequences of all 1's.  Thus, they are disjoint for $1 \leq t < n, t \in \mathbb{N}.$
\end{IEEEproof}

\section{Observations}

	For either insertions or deletions, the most practical value of $t$ to first consider is $ \floor{ \frac{d_{min}}{2}}+1$ i.e. one greater than the guaranteed number of correctable insertions/deletions.  In this section we discuss the observed behavior of various codes for a small number of extra insertions and raise some open questions. 
		
	Recall the single insertion correcting VT-code.  Given a VT-code of length $n,$ it is known that $a=0$ always gives the highest cardinality, though there are often multiple values of $a$ that give rise to VT-codes of maximum cardinality. 
	
	Among $a$ values that give maximum cardinality, we can select the one which maximizes $W_K(VT_a(n)),$ or $U_K(VT_a(n))$ for $K \in \{ \text{USC}_{t},  \text{UIC}_{t}, \text{UDC}_{t}, \text{UBC}_{t}\},$ for some $t > 1$ of our interest.  In the case of $t=2$ insertions for example, decoding is efficient. Given any algorithm for decoding $1$ insertion that outputs a subsequence of the received word (e.g. using the general approach given in \cite{Lev2}), $2$ insertions can be decoded by simply running that decoding algorithm on each of the $1$-subsequences of the received word and checking whether the decoding is a codeword.

	We observed that by varying $a,$ $W_K(VT_a(n))$ and $U_K(VT_a(n))$ for $K \in \{ \text{USC}_{2},  \text{UIC}_{2}, \text{UDC}_{2}, \text{UBC}_{2}\}$ can change significantly for smaller $n$, making this a practical observation for relatively small $n.$  However, as $n$ increases, we have observed that $W_K(VT_a(n))$ and $U_K(VT_a(n))$ approach $0$ for $K \in \{ \text{USC}_{2},  \text{UIC}_{2}, \text{UDC}_{2}, \text{UBC}_{2}\}.$  Through our observations, we ask the following question about the $\text{USC}_{2},  \text{UIC}_{2},$ $\text{UDC}_{2},$ and  $\text{UBC}_{2}.$
	
	
	For an optimal length $n$ code $C$ with minimum Levenshtein distance $d_{\text{min}} = \Theta(1),$ is it always true that $\lim_{n \to \infty} U_{K}(C) = 0$ and $\lim_{n \to \infty} W_{K}(C) = 0$ for $K \in \{ \text{USC}_{\floor{\frac{d_{\text{min}}}{2}}+1},  \text{UIC}_{\floor{\frac{d_{\text{min}}}{2}}+1}, \text{UDC}_{\floor{\frac{d_{\text{min}}}{2}}+1}, \text{UBC}_{\floor{\frac{d_{\text{min}}}{2}}+1}\}$?

	On the other hand, for the code $\\ C = \{ \underbrace{00\ldots 0}_{n \text{ 0's}}, \quad \underbrace{00\ldots 0}_{\frac{n}{2} \text{ 0's}} \underbrace{11\ldots 1}_{\frac{n}{2} \text{ 1's}}, \quad  \underbrace{11\ldots 1}_{\frac{n}{2} \text{ 1's}} \underbrace{00\ldots 0}_{\frac{n}{2} \text{ 0's}}, \quad \underbrace{11\ldots 1}_{n \text{ 1's}}\}, $ we proved that $\lim_{n \to \infty} U_{K}(C) = 1$ and $\lim_{n \to \infty} W_{K}(C) = 1$ for $K \in \{ \text{USC}_{\frac{n}{2}}, \text{UIC}_{\frac{n}{2}} \}.$  	The code $C = \{0_n, 1_n\}$ is an example where for  $\lim_{n \to \infty} U_{K}(C) = 1$ and $\lim_{n \to \infty} W_{K}(C) = 1$ for $K = \text{USC}_{\floor{\frac{d_{\text{min}}}{2}}+b}, \text{UIC}_{\floor{\frac{d_{\text{min}}}{2}}+b} $ for constant $b \in \mathbb{N}$ (see Lemmas \ref{lem:lim_n}, \ref{lem:UIC_lim_n}).  Through these observations, we ask the following question.

	For a length $n$ code $C$ with minimum Levenshtein distance $d_{\text{min}} = \Theta(n),$ is it always true that $\lim_{n \to \infty} U_{K}(C) = 1$ and $\lim_{n \to \infty} W_{K}(C) = 1$ for $K \in \{ \text{USC}_{\floor{\frac{d_{\text{min}}}{2}}+b},  \text{UIC}_{\floor{\frac{d_{\text{min}}}{2}}+b}\}$ and constant $b\in \mathbb{N}?$

	We are also interested in the analogues of these questions when $d_{\text{min}}$ and $b$ are different functions of $n.$  

\section{Conclusion}

In this work, we examined the unique decoding capability of codes from insertions and deletions beyond the guaranteed number of correctable insertions/deletions.  We defined several probabilistic channels  as a framework to understand what happens when $t$ insertions or $t$ deletions occur, regardless of whether the real-life channel always makes $t$ insertions or $t$ deletions.  We computed tight upper bounds on the probability of unique decoding for the channels, and studied the limiting behavior of the bounds. We then found upper bounds specific to particular classes of codes, such as linear codes and VT-codes.  We also studied the positivity of the measures, and raised several open questions.

%

\section{Appendix}

\subsection{Proof of Theorem~\ref{lem:Main}}

\textbf{Theorem 1.} Let $n_1,n_2,t_1,t_2 \in \mathbb{N}$ and $t_1 + n_1 = t_2 + n_2$.  Then, we have
	
	\begin{align} \label{eq:N_dots} \ddot{N}^+_q(n_1, n_2, t_1, t_2) =  \sum_{k = n_1}^{t_2} \sum_{i=0}^{k-n_1} \binom{k}{i}(q-2)^i \binom{n_2 + t_2}{k}. \!
	\end{align}

\begin{IEEEproof}
	First, we will show that $|I_{t_1}(0_{n_1}) \cap I_{t_2}(1_{n_2})| = \sum_{k = n_1}^{t_2} \sum_{i=0}^{k-n_1} \binom{k}{i}(q-2)^i \binom{n_2 + t_2}{k} $ for all $n_1,n_2,t_1,t_2,q \in \mathbb{N}$ where $t_1 + n_1 = t_2 + n_2$.   We will then prove that \[\min_{\substack{X \in F_q^{n_1}, Y \in F_q^{n_2}}} |I_{t_1}(X) \cap I_{t_2}(Y)| =  |I_{t_1}(0_{n_1}) \cap I_{t_2}(1_{n_2})|\] by strong induction on $N = n_1 + t_1 = n_2 + t_2$.

	Without loss of generality, suppose that $n_1  \leq n_2$.  We observe that for a sequence to be in the intersection of $I_{t_1}(0_{n_1})$ and $I_{t_2}(1_{n_2})$, the sequence must contain at least $n_1$ zeros and $n_2$ ones.  Thus if $t_2 < n_1$, we have that $|I_{t_1}(0_{n_1}) \cap I_{t_2}(1_{n_2})| = 0.$  
	
	Let $N = n_1 + t_1 = n_2 + t_2$. If $t_2 \geq n_1$, we observe that $I_{t_1}(0_{n_1}) \cap I_{t_2}(1_{n_2})$ consists of all length $N$ sequences containing $N-k$ ones and $p$ zeros where $n_1 \leq k \leq N-n_2$ and $ n_1 \leq p \leq k.$   
	
	To count the number of length $N$ sequences with $N-k$ ones, we first observe that the set $S_{0_{n_1}}$ containing all $k$-supersequences of $0_{n_1}$ that contain no ones has cardinality equal to $I_{k-n_1}(n_1, q-1)$. Let $A_k(T)$ be the set of all sequences formed by inserting $k$ ones into a sequence $T$.  Consider $T_1, T_2 \in S_{0_{n_1}}, T_1 \neq T_2$.  Then clearly $|A_k(T_1) \cap A_k(T_2)| = 0$.  To count the number of sequences in $A_k(T_1)$, we can view each of the $k+1$ spaces adjacent to sequence elements in $T_1$ as a labeled bin, and the $N-k$ ones that will be inserted into the sequence as $N-k$ unlabeled balls that are being tossed into the $k+1$ bins.
	
	Using the formula, for unlabeled balls tossed into labeled bins, we see that $|A_k(T_1)| = \binom{(N-k)+(k+1)-1}{(k+1)-1} = \binom{N}{k} = \binom{n_2 + t_2}{k}$.
	
	Thus, the number of supersequences in $I_{t_1}(0_{n_1}) \cap I_{t_2}(1_{n_2})$ with $N-k$ ones is equal to   $I_{k-n_1}(n_1,q-1)\binom{n_2 + t_2}{k}$
	
	Summing over the possible values of $k$, we obtain \begin{align*} &|I_{t_1}(0_{n_1}) \cap I_{t_2}(1_{n_2})| \\ 
	& = \sum_{k = n_1}^{t_2} I_{k-n_1}(n,q-1)\binom{n_2 + t_2}{k}\\
	& = \sum_{k = n_1}^{t_2} \sum_{i=0}^{k-n_1} \binom{n_1 + (k - n_1)}{i}(q-2)^i \binom{n_2 + t_2}{k}	 \\
	& = \sum_{k = n_1}^{t_2} \sum_{i=0}^{k-n_1} \binom{k}{i}(q-2)^i \binom{n_2 + t_2}{k}
	\end{align*}
	
	We proceed to show that \[\min_{\substack{X \in F_q^{n_1}, Y \in F_q^{n_2}}} |I_{t_1}(X) \cap I_{t_2}(Y)| =  |I_{t_1}(0_{n_1}) \cap I_{t_2}(1_{n_2})|\] by induction on $N = n_1 + t_1 = n_2 + t_2$. 
	
	The property trivially holds for $N = 0$.  For $N = 1$, $|I_{0}(0) \cap I_{0}(1)| = 0$ while in any other case, the cardinality must be $\geq 0$ since set cardinalities are always non-negative.
	
	Suppose the property holds for all $N$ in the range $ 0 \leq N \leq T$.  For $N = T+1$, we consider the pair of sequences $X = 0_{n_1}$ and $Y = 1_{n_2}$ for any $n_1, n_2, t_1, t_2 \in \mathbb{N}$ such that $t_1 + n_1 = t_2 + n_2 = T+1$.  From Lemma \ref{levTrick}, we see that
	
	\begin{align*} 
	|&I_{t_1}(0_{n_1}) \cap I_{t_2}(1_{n_2})| = \nonumber  \\
	&\quad|I_{t_1}(0_{n_1-1}) \cap I_{t_2-1}(1_{n_2})| + |I_{t_1-1}(0_{n_1}) \cap I_{t_2}(1_{n_2-1})|  \\ 
	&\quad+ (q-2)|I_{t_1-1}(0_{n_1}) \cap I_{t_2-1}(1_{n_2})|. 
	\end{align*}
	
	Now consider two arbitrary sequences $X = aX' \in F_q^{n_1}$ and $Y = bY' \in  F_q^{n_2}$.  We see from Lemma \ref{levTrick} that if $a \neq b$, then 
	
	\begin{align*} 
	|&I_{t_1}(X) \cap I_{t_2}(Y)| = \nonumber  \\
	&\quad|I_{t_1}(X') \cap I_{t_2-1}(bY')| + |I_{t_1-1}(aX') \cap I_{t_2}(Y')| \\
	&\quad + (q-2)|I_{t_1-1}(aX') \cap I_{t_2-1}(bY')|. 
	\end{align*}
	
	By the inductive hypothesis, \[|I_{t_1}(0_{n_1-1}) \cap I_{t_2-1}(1_{n_2})| \leq |I_{t_1}(X') \cap I_{t_2-1}(bY')|\] and \[|I_{t_1-1}(0_{n_1}) \cap I_{t_2}(1_{n_2-1})| \leq |I_{t_1-1}(aX') \cap I_{t_2}(Y')|\] and
	\[|I_{t_1-1}(0_{n_1}) \cap I_{t_2-1}(1_{n_2})| \leq |I_{t_1-1}(aX') \cap I_{t_2-1}(bY')|\].  
	
	Thus, if $a \neq b$, 
	
	\[ |I_{t_1}(0_{n_1}) \cap I_{t_2}(1_{n_2})| \leq |I_{t_1}(X) \cap I_{t_2}(Y)| .\]
	
	Now consider the case where $a = b.$  In this case, we have  
	\begin{align*} 
	|&I_{t_1}(X) \cap I_{t_2}(Y)| = \nonumber  \\
	&\quad|I_{t_1}(X') \cap I_{t_2}(Y')| + (q-1)|I_{t_1-1}(aX') \cap I_{t_2-1}(bY')|. 
	\end{align*}
	
	By the inductive hypothesis, we have 	\[|I_{t_1}(0_{n_1-1}) \cap I_{t_2}(1_{n_2-1})| \leq |I_{t_1}(X') \cap I_{t_2}(Y')|\] and \[|I_{t_1-1}(0_{n_1}) \cap I_{t_2-1}(1_{n_2})| \leq |I_{t_1-1}(aX') \cap I_{t_2-1}(bY')|\].
	
	We thus have that \begin{align*} |I&_{t_1}(X) \cap I_{t_2}(Y)| \\
	& \geq |I_{t_1}(0_{n_1-1}) \cap I_{t_2}(1_{n_2-1})| \\
	& + (q-1)|I_{t_1-1}(0_{n_1}) \cap I_{t_2-1}(1_{n_2})|
	\end{align*}
	
	We will complete the proof by showing that 	\begin{align*}
	|I&_{t_1}(0_{n_1-1}) \cap I_{t_2}(1_{n_2-1})| + (q-1)|I_{t_1-1}(0_{n_1}) \cap I_{t_2-1}(1_{n_2})| \\
	&= |I_{t_1}(0_{n_1-1}) \cap I_{t_2}(1_{n_2-1})| + |I_{t_1-1}(0_{n_1}) \cap I_{t_2-1}(1_{n_2}) \\
	&+  (q-2)|I_{t_1-1}(0_{n_1}) \cap I_{t_2-1}(1_{n_2})| \\
	&\geq |I_{t_1-1}(0_{n_1}) \cap I_{t_2}(1_{n_2-1})| + |I_{t_1}(0_{n_1-1}) \cap I_{t_2-1}(1_{n_2})| \\ 
	&+ (q-2)|I_{t_1-1}(0_{n_1}) \cap I_{t_2-1}(1_{n_2})| \\
	& = |I_{t_1}(0_{n_1}) \cap I_{t_2}(1_{n_2})|
	\end{align*}
	
	To show the above, it suffices to prove that 
	\begin{align*}
	&|I_{t_1-1}(0_{n_1}) \cap I_{t_2-1}(1_{n_2}) + |I_{t_1}(0_{n_1-1}) \cap I_{t_2}(1_{n_2-1})| \\
	&\geq |I_{t_1}(0_{n_1-1}) \cap I_{t_2-1}(1_{n_2})| +  |I_{t_1-1}(0_{n_1}) \cap I_{t_2}(1_{n_2-1})|
	\end{align*}
	
	We begin by manipulating the left side of the inequality.
	
	\begin{align*} &|I_{t_1-1}(0_{n_1}) \cap I_{t_2-1}(1_{n_2})|  
	+ |I_{t_1}(0_{n_1-1}) \cap I_{t_2}(1_{n_2-1})| \\
	& = \sum_{k = n_1}^{t_2-1} \sum_{i=0}^{k-n_1} \binom{k}{i}(q-2)^i \binom{n_2 + t_2-1}{k} \\
	& +  \sum_{k = n_1-1}^{t_2} \sum_{i=0}^{k-n_1+1} \binom{k}{i}(q-2)^i \binom{n_2-1 + t_2}{k} \\ 
	& = \sum_{k = n_1}^{t_2-1} \sum_{i=0}^{k-n_1} (q-2)^i \bigg( \binom{k}{i}\binom{n_2-1 + t_2}{k} \\
	&+ \binom{k-1}{i}\binom{n_2-1 + t_2}{k-1}\bigg) \\
	&+ \sum_{k=t_2-1}^{t_2} \sum_{i=0}^{k-n_1+1} \binom{k}{i} (q-2)^i \binom{n_1+t_1-1}{k} 
	\end{align*}
	
	The second equality above follows from splitting up the second term.  This expression gives
	
	\begin{align*}
	& C_1 + \sum_{i=0}^{t_2-n_1} \binom{t_2-1}{i} (q-2)^i \binom{n_1+t_1-1}{t_2-1} \\
	&+ \sum_{i=0}^{t_2-n_1+1} \binom{t_2}{i} (q-2)^i \binom{n_1+t_1-1}{t_2} \\
	& = C_1 + C_2 + \sum_{i=0}^{t_2-n_1+1} (q-2)^i \binom{t_2}{i}\binom{n_2 + t_2 - 1}{t_2}
	\end{align*}
	where 
	\begin{align*}
	C_1 = &\sum_{k = n_1}^{t_2-1} \sum_{i=0}^{k-n_1} (q-2)^i \bigg( \binom{k}{i}\binom{n_2-1 + t_2}{k} \\
	&+ \binom{k-1}{i}\binom{n_2-1 + t_2}{k-1}\bigg)
	\end{align*}
	and \begin{align*}
	C_2 = \sum_{i=0}^{t_2-n_1} \binom{t_2-1}{i} (q-2)^i \binom{n_1+t_1-1}{t_2-1}.
	\end{align*}

	We now manipulate the right side of the inequality.
	
	\begin{align*}
	&|I_{t_1}(0_{n_1-1}) \cap I_{t_2-1}(1_{n_2})| + |I_{t_1-1}(0_{n_1}) \cap I_{t_2}(1_{n_2-1})|  \\ 
	&= \sum_{k = n_1-1}^{t_2-1} \sum_{i=0}^{k-n_1+1} \binom{k}{i}(q-2)^i \binom{n_2 + t_2-1}{k} \\
	& + \sum_{k = n_1}^{t_2} \sum_{i=0}^{k-n_1} \binom{k}{i}(q-2)^i \binom{n_2 + t_2-1}{k} \\
	& = \sum_{k = n_1}^{t_2} \sum_{i=0}^{k-n_1} (q-2)^i \bigg( \binom{k}{i}\binom{n_2-1 + t_2}{k} \\
	&+ \binom{k-1}{i}\binom{n_2-1 + t_2}{k-1}\bigg) \\
	& = \sum_{k = n_1}^{t_2-1} \sum_{i=0}^{k-n_1} (q-2)^i \bigg( \binom{k}{i}\binom{n_2-1 + t_2}{k} \\
	&+ \binom{k-1}{i}\binom{n_2-1 + t_2}{k-1}\bigg) \\	
	& + \sum_{i=0}^{t_2-n_1} (q-2)^i \bigg( \binom{t_2}{i}\binom{n_2-1 + t_2}{t_2} \\
	&+ \binom{t_2-1}{i}\binom{n_2-1 + t_2}{t_2-1}\bigg) \\
	& = C_1 + \sum_{i=0}^{t_2-n_1} (q-2)^i \bigg( \binom{t_2}{i}\binom{n_2-1 + t_2}{t_2} \\
	&+ \binom{t_2-1}{i}\binom{n_2-1 + t_2}{t_2-1}\bigg) \\
	& = C_1 +\sum_{i=0}^{t_2-n_1} (q-2)^i \binom{t_2}{i}\binom{n_2-1 + t_2}{t_2} \\ 
	& + \sum_{i=0}^{t_2-n_1} (q-2)^i  \binom{t_2-1}{i}\binom{n_2-1 + t_2}{t_2-1} \\
	& = C_1 + C_2 + \sum_{i=0}^{t_2-n_1} (q-2)^i \binom{t_2}{i}\binom{n_2 + t_2 - 1}{t_2}
	\end{align*}
	Clearly,
	\begin{align*}
	\sum_{i=0}^{t_2-n_1+1} &(q-2)^i \binom{t_2}{i}\binom{n_2 + t_2 - 1}{t_2} \\
	& \geq \sum_{i=0}^{t_2-n_1} (q-2)^i \binom{t_2}{i}\binom{n_2 + t_2 - 1}{t_2} 
	\end{align*}

	Thus, we have

	\begin{align*}
	&|I_{t_1-1}(0_{n_1}) \cap I_{t_2-1}(1_{n_2}) + |I_{t_1}(0_{n_1-1}) \cap I_{t_2}(1_{n_2-1})| \\
	&\geq |I_{t_1}(0_{n_1-1}) \cap I_{t_2-1}(1_{n_2})| +  |I_{t_1-1}(0_{n_1}) \cap I_{t_2}(1_{n_2-1})|
	\end{align*}
	
	This proves,
	\begin{align*}
	|I&_{t_1}(0_{n_1-1}) \cap I_{t_2}(1_{n_2-1})| + (q-1)|I_{t_1-1}(0_{n_1}) \cap I_{t_2-1}(1_{n_2})| \\
	&\geq |I_{t_1-1}(0_{n_1}) \cap I_{t_2}(0_{n_2-1})| + |I_{t_1}(0_{n_1-1}) \cap I_{t_2-1}(0_{n_2})| \\ 
	&+ (q-2)|I_{t_1-1}(0_{n_1}) \cap I_{t_2-1}(0_{n_2})| 
	\end{align*}
	
	Thus, if $a = b$, we have 
	
	\[ |I_{t_1}(0_{n_1}) \cap I_{t_2}(1_{n_2})| \leq |I_{t_1}(X) \cap I_{t_2}(Y)| \]
	
	and we have proved the inductive step.
\end{IEEEproof}

\subsection{Proof of Lemma \ref{lem:max_pairs_min_intersection}}

Lemma \ref{lem:max_pairs_min_intersection} is restated below, and follows directly form the proceeding auxiliary lemmas along with Theorem \ref{lem:Main}.

	\textbf{Lemma \ref{lem:max_pairs_min_intersection}.} Let $n_1,n_2,t_1,t_2 \in \mathbb{N}$ and $N = t_1 + n_1 = t_2 + n_2$.  Then for any length $n_1$ sequence $X$ and any length $n_2$ sequence $Y$ such that $d_L(X, Y) = n_1 + n_2$, we have that 
	
	\begin{align*} &|I_{t_1}(X) \cap I_{t_2}Y)| \\
	& = \ddot{N}^+_q(n_1, n_2, t_1, t_2) =  \sum_{k = n_1}^{t_2} \sum_{i=0}^{k-n_1} \binom{k}{i}(q-2)^i \binom{n_2 + t_2}{k}.
	\end{align*} 

\begin{lemma} \label{lem:MaxDistPair}
	Let $n_1,n_2,t_1,t_2 \in \mathbb{N}$ and $N = t_1 + n_1 = t_2 + n_2$.  Then for any length $n_1$ sequence $X$ and any length $n_2$ sequence $Y$ such that $d_L(X, Y) = n_1 + n_2$, we have that $Y$ only contains elements from some subset $B$ of the alphabet $\mathbb{F}_{q}$, and $X$ only contains elements from the subset $\mathbb{F}_{q} \setminus B$.  Letting $q_B = |B|$, we have that
	
	\begin{align*} &|I_{t_1}(X) \cap I_{t_2}(Y)| \\
	&= \sum_{k = 0}^{N-n_1-n_2} I_k(n_1, q-q_B) I_{N-k-n_1-n_2} (n_2, q_B) \binom{N}{t_1 - k}.
	\end{align*}
	
\end{lemma}

\begin{IEEEproof}
	Clearly, if $d_L(X, Y) = n_1 + n_2$, we have that $X$ and $Y$ do not have any elements in common.  Thus we consider the set of elements in $Y$ as a new alphabet $B$, which is clearly a subset of the alphabet $\mathbb{F}_{q}$.  We then consider the subset  $\mathbb{F}_{q} \setminus B$ as the alphabet for $X$.  
	
	To count the number of sequences in $I_{t_1}(X) \cap I_{t_2}(Y)$, we partition the sequences based on the number $n_1 + k$ of elements from  $\mathbb{F}_{q} \setminus B$ that the sequence contains.  Given some value of $k$, there are $I_k(n_1, q-q_B)$ possibilities for the sequences's subsequence of elements from  $\mathbb{F}_{q} \setminus B$.
	
	A sequence in $I_{t_1}(X) \cap I_{t_2}(Y)$ with $k$ additional elements from  $\mathbb{F}_{q} \setminus B$ must have $N - n_1 - k$ elements from $B$.   Given $k$, there are $I_{N-k-n_1-n_2} (n_2, q_B)$ possibilities for the sequence's subsequence of elements from $B$.  
	
	Given the sequence's length $n_1 + k$ subsequence $T_X$ of elements from  $\mathbb{F}_{q} \setminus B$ and the sequence's length $N - n_1 - k$ subsequence $T_Y$ of elements from $B$, the number of possible sequences containing both of these subsequences is the number of ways $T_Y$ can be interleaved with $T_X$.  This can be counted using a balls and bins approach.
	
	We consider the $n_1 + k + 1$ possible insertion positions into $T_X$ as labeled bins, and the $N-k-n_1$ positions of elements from $T_Y$ as unlabeled balls.  There are thus $\binom{(n_1 +k+1) + (N-k-n_1) -1}{N-k-n_1}$  ways for $T_Y$ to be interleaved with $T_X.$
	
	$k$ can range from $0$ to $N-n_1-n_2$ so we have
	\begin{align*} &|I_{t_1}(X) \cap I_{t_2}(Y)| \\
	&= \sum_{k = 0}^{N-n_1-n_2} I_k(n_1, q-q_B) I_{N-k-n_1-n_2} (n_2, q_B) \binom{N}{t_1 - k}. 
	\end{align*}
\end{IEEEproof}


\begin{lemma} \label{lem:0_1_intersection}
	Let $n_1,n_2,t_1,t_2 \in \mathbb{N}$ and $N = t_1 + n_1 = t_2 + n_2$.  Consider any subset  $B \subseteq \mathbb{F}_{q}$ such that $1 \leq q_B \leq q - 1$, where $q_B = |B|$.  Then, 
	
	\begin{align*} &|I_{t_1}(0_{n_1}) \cap I_{t_2}(1_{n_2})| \\
	&= \sum_{k = 0}^{N-n_1-n_2} I_k(n_1, q-q_B) I_{N-k-n_1-n_2} (n_2, q_B) \binom{N}{t_1 - k}.
	\end{align*} 
\end{lemma}

\begin{IEEEproof}
	Using a similar approach as in Lemma \ref{lem:MaxDistPair}, we will choose a sub-alphabet for the sequence $1_{n_2}$, and a sub-alphabet for $0_{n_1}$.  
	
	$0_{n_1}$ and $1_{n_2}$ have no elements in common.  So, the alphabet $A$ for $1_{n_2}$ can be chosen to be any subset of  $\mathbb{F}_{q}$ such that $1 \in A$, and $0 \notin A$, and $|A| = q_B$.  Once the alphabet for $1_{n_2}$ is chosen, the alphabet for $0_{n_1}$ will be $\mathbb{F}_{q} \setminus A$. Thus, we have that
	
	\begin{align*} &|I_{t_1}(0_{n_1}) \cap I_{t_2}(1_{n_2})| \\
	&= \sum_{k = 0}^{N-n_1-n_2} I_k(n_1, q-q_B) I_{N-k-n_1-n_2} (n_2, q_B) \binom{N}{t_1 - k}.
	\end{align*} 
	
	by following the proof in Lemma \ref{lem:MaxDistPair}  
\end{IEEEproof}

\end{document}